\documentstyle[12pt]{article}

\advance \textheight by \topskip
\begin{document}

\thispagestyle{empty}
\begin{flushright}
{\bf DFTUZ/95/25}\\
{\bf DFUPG-105/95}\\
{\bf hep-th/9511028}
\end{flushright}
$\ $
\vskip 2truecm

\begin{center}

{ \Large \bf Hidden Supersymmetry of a}\\
\vskip0.5cm
{\Large \bf $P,T-$Invariant $3D$ Fermion System}\\
\vskip1.0cm
{ \bf Mikhail Plyushchay${}^{a}${}\footnote{
On leave from {\it Institute for High Energy Physics, Protvino,
Moscow Region, Russia}; e-mail: mikhail@cc.unizar.es}
and
Pasquale Sodano${}^{b}${}\footnote{E-mail: sodano@perugia.infn.it}\\[0.3cm]
{\it ${}^{a}$Departamento de F\'{i}sica Te\'orica, Facultad de Ciencias}\\
{\it Universidad de Zaragoza, 50009 Zaragoza, Spain}\\
{\it ${}^{b}$Dipartimento di Fisica and Sezione I.N.F.N.}\\
{\it Universit\'a di Perugia, via A. Pascoli, I-06100 Perugia, Italy}}
\end{center}

\vskip2.0cm
\begin{center}
                            {\bf Abstract}
\end{center}
We show that a (2+1)-dimensional $P,T-$invariant free fermion
system, relevant to $P,T-$conserving models of high-$T_c$
superconductivity, has a U(1,1) dynamical symmetry as well as an
$N=3$ supersymmetry with  the even generator being a quadratic
function of the spin operator and of the generator of chiral
${\rm U_c(1)}$ transformations.  We demonstrate that the hidden
supersymmetry leads to a non-standard superextension of the
(2+1)-dimensional Poincar\'e group.  As a result, the one
particle states of the $P,T-$invariant fermion system realize an
irreducible representation of the Poincar\'e supergroup labelled
by the zero eigenvalue of the superspin operator.

\newpage

{\bf 1.}
Planar gauge field theories have many interesting theoretical features such
as gauge invariant local topological mass \cite{1,2},
fractional spin \cite{1,2,3,4}
and statistics \cite{5}.
The exotic spin and statistics of planar theories --- together with
the appearance of $P-$ and $T-$breaking mass terms in the Lagrangians of
the massive
Dirac spinor field and topologically massive vector
U(1) gauge field \cite{1,2} ---
are a consequence of the simple fact that in 2+1 dimensions
the spin is a pseudoscalar quantity.

Especially when constructing gauge models of
high-T${}_c$ superconductors \cite{6},
one is interested in having a $P-$ and
$T-$invariant  topologically massive vector gauge field
and a $P-$ and $T-$invariant massive Dirac spinor field.
For this purpose one usually introduces doublets of these fields
with mass terms having opposite signs \cite{6,7}.

In this paper we shall investigate the hidden symmetries of
the simplest $P-$ and $T-$invariant fermion theory , namely, we
shall investigate the model described by the Lagrangian
\begin{equation}
{\cal L}=\bar{\psi}_u(p\gamma+m)\psi_u
+\bar{\psi}_d(p\gamma-m)\psi_d.
\label{1}
\end{equation}
This model is invariant under a global
${\rm U_c(1)}$ symmetry describing chiral
rotations, to which is associated the conserved chiral
current $I_\mu=\frac{1}{2}(\bar{\psi}_u\gamma_\mu\psi_u-
\bar{\psi}_d\gamma_\mu\psi_d)$.
This global symmetry
is promoted to a local gauge symmetry in all the models of
refs. \cite{6,7}.

We shall demonstrate that the free fermion system described
by (\ref{1}) has not only the global
${\rm U_c(1)}$ symmetry,
but a broader U(1,1)=SU(1,1)$\times$U(1) symmetry as
well as a hidden N=3 supersymmetry. As we shall see,
these symmetries form a dynamical group symmetry of the
one particle states, i.e. the quantum mechanical states, of
the free fermion theory described
by (\ref{1}). Furthermore,
the one particle
states realize an irreducible representation of a
nonstandard Poincar\'e supergroup,
whose generators we shall construct explicitly. Therefore, the
dynamical symmetries of the
$P,T-$invariant fermion model described by (\ref{1})
differ substantially
from those of its $P,T-$non-invariant counterpart.

$\ $

{\bf 2.}
To begin, let us introduce the spinor function
$\Psi$,
$\Psi^t=(\psi^t_u,\psi^t_d)$,
in terms of which the Lagrangian
(\ref{1}) is represented as ${\cal L}=\bar{\Psi}{\cal D}\Psi$,
with
\begin{equation}
{\cal D}=p\gamma\otimes 1+m\cdot 1\otimes\sigma_3.
\label{2}
\end{equation}
In (\ref{2}) we use the (2+1)-dimensional $\gamma$-matrices realized
via the Pauli matrices, $\gamma^0=\sigma_3$,
$\gamma^i=i\sigma_i$, $i=1,2$, which satisfy the relation
$\gamma_\mu\gamma_\nu=-\eta_{\mu\nu}+
i\epsilon_{\mu\nu\lambda}\gamma^\lambda$ with
metric $\eta_{\mu\nu}=diag(-,+,+)$ and
completely antisymmetric tensor $\epsilon^{\mu\nu\lambda}$
normalized as $\epsilon^{012}=1$.
To avoid a cumbersome notation we do not write explicitly
the spinor indices on which the matrices $\gamma_\mu$ act.
The second factor in the direct product in
(\ref{2}), being $1$ or $\sigma_3$ (and $\sigma_{1,2}$ in
other operators below), acts in the two-dimensional space with
indices $u=1$ and $d=2$.  With this notation
the equations of motion are
\begin{equation}
{\cal D}\Psi=0.
\label{3}
\end{equation}
The
operator ${\cal D}$ satisfies the relation
\begin{equation}
{\cal D}^2=-K+4m{\cal N}{\cal D},
\label{4}
\end{equation}
where
$K=p^2+m^2$ is the Klein-Gordon operator and
\begin{equation}
{\cal N}=\frac{1}{2}\cdot 1\otimes\sigma_3.
\label{5}
\end{equation}
The operators commuting with
(\ref{2}) are called physical operators or integrals of motion
since they are symmetry generators of the
quantum mechanical system
\cite{8}. The operator ${\cal N}$ is then the integral of motion
generating the global chiral rotations
$\Psi\rightarrow \Psi'=\exp(ig{\cal N})\Psi$, where $g$ is a
constant transformation parameter.  Another integral of motion
is the spin operator
\begin{equation}
{\cal S}=-\frac{1}{2}\gamma^{(0)}\otimes 1.
\label{6}
\end{equation}
In the following we shall
work in the momentum representation.
In eq.  (\ref{6}) we have introduced the notation
$\gamma^{(0)}=\gamma^\mu e^{(0)}_\mu$,
$e^{(0)}_\mu=p_\mu/\sqrt{-p^2}$, taking into account that, due to (\ref{4}),
$-p^2\neq 0$ on the physical subspace
(\ref{3}).
We add to the
time-like unit vector $e^{(0)}_\mu$ the space-like quantities
$e^{(i)}_\mu=e^{(i)}_\mu(p)$, $i=1,2$, in order to have
a complete oriented triad $e^{(\alpha)}_\mu$, $\alpha=0,1,2$,
\begin{equation} e^{(\alpha)}_\mu
\eta_{\alpha\beta}e^{(\beta)}_\nu=\eta_{\mu\nu},\quad
\epsilon_{\mu\nu\lambda}e^{(0)\mu}e^{(1)\nu}e^{(2)\lambda}=1.
\label{7}
\end{equation}
Note that  $e^{(i)}_\mu$, $i=1,2,$ are not
Lorentz vectors; however, their explicit form is not needed
for our derivation (see, e.g., ref. \cite{9}). We shall
use also the notation $\gamma^{(\alpha)}=\gamma^\mu
e^{(\alpha)}_\mu$.

Let us consider
the following mutually
conjugate operators ${\cal Q}^\pm$,
\begin{equation}
{\cal Q}^{\pm}=({\cal
Q}^\mp)^\dagger=\pm\frac{i}{8}\left(\gamma^{(1)}\mp
i\gamma^{(2)}\right)\otimes(\sigma_1 \pm i\sigma_2).
\label{8}
\end{equation}
In (\ref{8}) the conjugation is with respect to the internal
scalar product
$(\Psi_1,\Psi_2)=\bar{\Psi}_1\Psi_2$. This is an
indefinite scalar product due to the presence of the
$\gamma^0\otimes 1$ factor.
The operators defined in (\ref{8}) satisfy the
weak condition \cite{8}
\[
[{\cal Q}^\pm,{\cal D}]=\pm 2 {\cal Q}^\pm\cdot
\left(\sqrt{-p^2}-m\right)\approx 0
\]
on the physical subspace defined by (\ref{3}),
implying that --- in addition  to
${\cal N}$ and ${\cal S}$ --- also
${\cal Q}^\pm$ must be regarded as integrals of motion.
One easily verifies that ${\cal N}$, ${\cal S}$ and ${\cal Q}^{\pm}$
satisfy the algebra
\[
[{\cal N},{\cal Q}^\pm]=\pm {\cal Q}^\pm,\quad
[{\cal S},{\cal Q}^\pm]=\pm {\cal Q}^\pm,\quad
[{\cal Q}^+,{\cal Q}^-]=-\frac{1}{4}({\cal N}+{\cal S}),\quad
[{\cal N},{\cal S}]=0.
\]
Upon considering the following linear
combinations,
\begin{equation}
{\cal Q}_0=
-\frac{1}{2}({\cal N}+{\cal S}),\quad {\cal Q}_1={\cal Q}^+
+{\cal Q}^-,\quad
{\quad Q}_2=i({\cal Q}^+ -{\cal Q}^-),
\label{9}
\end{equation}
we find
that the operators ${\cal Q}_\alpha$, $\alpha=0,1,2$, form
an $su(1,1)$ algebra:
\begin{equation}
[{\cal Q}_{\alpha},{\cal
Q}_\beta]=-i\epsilon_{\alpha\beta\gamma}{\cal Q}^\gamma.
\label{10}
\end{equation}
In (\ref{10}) ${\cal Q}^\alpha=\eta^{\alpha\beta}{\cal Q}_\beta$.
The operators ${\cal Q}_\alpha$,
together with the operator
\begin{equation}
{\cal U}=\frac{1}{2}({\cal S}-{\cal N}),
\label{11}
\end{equation}
which coincides (up to a
factor)  with the operator ${\cal D}$ on the physical subspace
(\ref{3}), are the generators of a
U(1,1)=SU(1,1)$\times$U(1) symmetry with the algebra given by
(\ref{10}) and $[{\cal U},{\cal Q}_\alpha]=0$.  The
Casimir operator of the $su(1,1)$ algebra is ${\cal
C}={\cal Q}_\alpha {\cal Q}^{\alpha}.$
Here it has the form
\begin{equation}
{\cal C}=-\frac{3}{8}(4{\cal NS}+1),
\label{12}
\end{equation}
and takes the value ${\cal C}=-3/4$ on the physical subspace
(\ref{3}).  Therefore, the equations of motion (\ref{3}) imply
that the physical subspace is
formed by the states with $-p^2=m^2$,
which are annihilated by the U(1) generator ${\cal U}$. On this
subspace the hidden SU(1,1) symmetry acts irreducibly.

On the other hand, since
${\cal Q}_1^2={\cal Q}_2^2=-{\cal Q}_0^2$,
the integrals (\ref{9}) together
with the Casimir operator (\ref{12}) form also
the following $s(3)$ \cite{10} superalgebra:
\begin{equation}
\{{\cal Q}_\alpha,{\cal
Q}_\beta\}=\eta_{\alpha\beta}\cdot\frac{2}{3}
{\cal C}, \qquad
[{\cal Q}_\alpha,{\cal C}]=0.
\label{13}
\end{equation}
Thus, the system described by (\ref{1}) has a hidden
$N=3$ dynamical supersymmetry.
In our construction, we treat the operator ${\cal D}$ as the
`hamiltonian' since we consider the operators commuting with it
as integrals of motion \cite{8}.  Since the even
generator of the superalgebra (\ref{13}) differs from the
`hamiltonian' ${\cal D}$, the $N=3$ hidden supersymmetry
turns out to be analogous to the hidden supersymmetry of a
3-dimensional monopole \cite{11}, where one of the even
generators, being the square of one of the two odd supercharges, is
also different from the corresponding hamiltonian.
The latter system
admits SU(1,1) as a subgroup
of its dynamical symmetry group \cite{11}.
Other systems showing similar hidden supersymmetries
have been analyzed in ref. \cite{12}.

The integrals ${\cal S}$ and ${\cal N}$ as well as their linear
combinations ${\cal Q}_0$ and ${\cal U}$  and the
quadratic operator (\ref{12}) have been written as
covariant operators, whereas
the integrals ${\cal Q}_i$ have not since
they are defined in terms of the noncovariant
quantities $\gamma^{(i)}=\gamma^\mu e^{(i)}_\mu$.
Nevertheless, one may introduce a vector operator
\[
\Gamma_\mu=-\frac{1}{4}(\eta_{\mu\nu}-\frac{p_\mu
p_\nu}{p^2})\gamma^\nu\otimes \sigma_2
-\frac{1}{4\sqrt{-p^2}}\epsilon_{\mu\nu\lambda}p^\nu\gamma^\lambda\otimes
\sigma_1,
\]
such that $\Gamma_\mu
e^{(i)\mu}={\cal Q}_i$. $\Gamma_\mu$ has
only two independent components since it is
transverse to $p_\mu$, $\Gamma_\mu p^\mu=0$. In terms of $\Gamma_\mu$
one may construct the Lorentz vector operator
\begin{equation}
\tilde{\cal Q}_\mu=\Gamma_\mu+e^{(0)}_\mu {\cal Q}_0
\label{14}
\end{equation}
so that
$\tilde{\cal Q}{}^{(\alpha)}={\cal
Q}^\alpha$, $\tilde{\cal
Q}_\mu \tilde{\cal Q}^\mu={\cal C}$.
As a consequence,
the operators defined in (\ref{14}) and (\ref{12})
satisfy the covariant
(anti)commutation relations generalizing
(\ref{10}), (\ref{13}).
With this notation, eqs. (\ref{10}) and (\ref{13})
may be written as
\begin{equation}
\tilde{\cal Q}_\mu\tilde{\cal
Q}_\nu= \eta_{\mu\nu}\cdot \frac{1}{3}{\cal
C}-\frac{i}{2}\epsilon_{\mu\nu\lambda}\tilde{\cal Q}{}^\lambda,
\qquad
[\tilde{\cal Q}_\mu,{\cal C}]=0.
\label{16}
\end{equation}
Eqs. (\ref{16}) together with the commutation relations
\begin{equation}
[\tilde{\cal Q}_\mu,{\cal U}]=[{\cal U},{\cal C}]=0
\label{17}
\end{equation}
covariantly manifests the dynamical (super)symmetry
of the system.

$\ $

{\bf 3.}
It is interesting to observe
that the hidden dynamical symmetry yields a
nonstandard super-extension of the Poincar\'e group.
The generators of the superextended Poincar\'e group
are $J_\mu$, $p_\mu$ and $\tilde{\cal Q}_\mu$, where $J_\mu$
is the total angular momentum operator,
\[
J_\mu=x_\mu p_\nu-x_\nu p_\mu
-\frac{1}{2}\gamma_\mu\otimes 1,
\]
and the vector generator $\tilde{\cal Q}_\mu$,
\[
[J_\mu,\tilde{\cal Q}_\nu]=-i\epsilon_{\mu\nu\epsilon}
\tilde{\cal Q}^\lambda,
\]
is the odd generator satisfying the anticommutation
relations given by eq. (\ref{16}). Since it commutes with
$p_\mu$, $[\tilde{\cal Q}_\mu,p_\nu]=0$, one of the
Casimir operators of the Poincar\'e supergroup
is $p^2$. In order to find the second Casimir operator
and clarify its meaning,
we consider the vector operator
\[
{\cal J}_\mu=J_\mu-\tilde{\cal Q}_\mu.
\]
It satisfies the same (2+1)-dimensional Lorentz algebra
(or $su(1,1)$ algebra) as the generators $J_\mu$ and
$\tilde{\cal Q}_\mu$ themselves,
\[
[{\cal J}_\mu,{\cal J}_\nu]=-i\epsilon_{\mu\nu\lambda}
{\cal J}^{\lambda},
\]
and, moreover, commutes with $\tilde{\cal Q}_\mu$.
Thus, the second Casimir operator is
\[
\tilde{\cal S}={\cal J}^\mu e^{(0)}_\mu.
\]
The operator $\tilde{\cal S}$
is the {\it superspin}.
Taking into account the explicit form of the operator
$\tilde{\cal Q}^{(0)}=-Q_0$ given by
eq. (\ref{9}), and the definition of spin,
$J^{(0)}={\cal S}$, one gets:
\[
\tilde{\cal S}={\cal U}=\frac{1}{2}({\cal S}-{\cal N}).
\]
Therefore, the operator ${\cal U}$ is the second
Casimir operator of the Poincar\'e supergroup, and it has the
meaning of a superspin.
One may easily verify that the
operator ${\cal C}$ is a quadratic
function of the superspin,
\begin{equation}
{\cal C}{}=3\tilde{\cal S}^2-\frac{3}{4}.
\label{17*}
\end{equation}
With the help of the explicit form of the spin operator ${\cal S}$
and of the generator of the chiral U${}_c$(1) transformations, ${\cal N}$,
one easily
finds that the spectrum of eigenvalues
of the superspin $\tilde{\cal S}$ is given by the series of numbers
$(-1/2,0,0,1/2).$
Thus, the
equations of motion
(\ref{3}) single out the physical states as the
subspace of states with
zero superspin.
These states satisfy also the Klein-Gordon equation.

$\ $

{\bf 4.}
We have displayed as a dynamical symmetry group
the hidden U(1,1)=
SU(1,1)$\times$U(1)
symmetry as well as the hidden N=3 supersymmetry
of the planar
$P,T-$invariant free fermion
model described by the Lagrangian (\ref{1}).
The N=3
supersymmetry turns out to be analogous to the hidden
supersymmetry of a 3-dimensional monopole since
the fermionic generators of the hidden supersymmetry
($\tilde{\cal Q}_\mu$) are the `square root' of
a bosonic integral
of motion ($\frac{1}{3}{\cal C}$) other than the `hamiltonian' (${\cal D}$)
\cite{11}. One may regard
the superspin
$\tilde{\cal S}$
as the Hamiltonian ${\cal H}$ for all
the matrix operator variables $\gamma^\mu\otimes 1$ (or $\gamma^{(\alpha)}
\otimes 1$) and $1\otimes\sigma_a$, $a=1,2,3$.
This can be easily seen \cite{13} from the pseudoclassical
model \cite{14}. Due to eq. (\ref{17*}),
the even generator of the
$N=3$ supersymmetry, $\tilde{\cal H}
\equiv\frac{2}{3}{\cal C}$, has the form
\begin{equation}
\tilde{\cal H}={\cal H}^2+b,
\label{18}
\end{equation}
with ${\cal H}=\sqrt{2}\tilde{\cal S}$ and $b=-1/2$.
As it is known, a supersymmetry with an
even generator of the form (\ref{18}) (with $b=0$)
takes place also in the description of
a relativistic electron in a uniform magnetic
field \cite{15} and in the analysis of
supersymmetric quantum mechanical
systems for which the anticommutator
of odd generators is a polynomial function of the hamiltonian
\cite{16}.

A $P,T-$non-invariant system of two
free fermion fields with mass terms of one sign
has a U(2)=SU(2)$\times$U(1) symmetry. The generators of the
SU(2) symmetry are $\frac{1}{2}\cdot 1\otimes \sigma_a$, $a=1,2,3$, whereas
the U(1) symmetry is generated by
the operator $\gamma^{(0)}\otimes 1+1\otimes 1$.
At first sight, one may easily conjecture that
--- in order to properly describe the dynamical
symmetries of a
$P,T-$invariant planar free fermion system ---
one should transform the generators of SU(2) into
the corresponding
generators of SU(1,1) with the simple trick
of changing one of them through
multiplication by $i$ (see, e.g., ref. \cite{4}).
This is not a correct procedure for the model investigated
in this letter since the
SU(1,1) generators constructed by means
of this trick are not
integrals of motion of the free fermion model described by (\ref{3}).
Furthermore,
the SU(2) generators of a $P,T-$non-invariant planar free fermion model
are Poincar\'e-invariant generators
of a merely internal symmetry, while the
SU(1,1) generators obtained with our construction
form a translationally-invariant Lorentz vector.
Our analysis clarifies an important property of the integrals of motion
associated with the Lagrangian (\ref{1}):
namely, it shows that the generators of SU(1,1),
being simultaneously odd generators of
an $N=3$ supersymmetry, can be  combined with the
Poincar\'e generators.
This results in the non-standard superextension
of the Poincar\'e group with a vector supercharge
explicitly exhibited in this letter.
The one particle states of the system described by (\ref{1})
realize an irreducible representation of the
Poincar\'e supergroup, labelled by the zero
eigenvalue of the superspin operator.

It would be interesting to look also for hidden
symmetries in the $P,T-$invariant system
of two free topologically massive vector U(1) gauge fields,
as well as for those arising in the
$P,T-$invariant system of two massive Dirac fields interacting
with a doublet of topologically massive gauge fields
\cite{1,2}. Furthermore,  it is intriguing to ascertain what
happens to the hidden supersymmetry of the
free fermion model described by (\ref{1}) if one switches on
the interaction with an external U(1) gauge field
${\cal A}_\mu$ coupled to the conserved  chiral
current $I_\mu=\bar{\Psi}{\cal N}_\mu \Psi$ \cite{13}.
We hope to address these problems
--- as well as the possibility of having a similar non-standard
Poincar\'e supergroup in (3+1)-dimensional space-time ---
in future publications.

$\ $

The research of M.P. was supported by MEC-DGICYT (Spain).
He also thanks I.N.F.N. for financial support and
the University of Perugia for hospitality.

\end{document}